\documentstyle[12pt]{article}

\def\hybrid{\topmargin -20pt	\oddsidemargin 0pt
	\headheight 0pt	\headsep 0pt
	\textwidth 6.25in	
	\textheight 9.5in	
	\marginparwidth .875in
	\parskip 5pt plus 1pt	\jot = 1.5ex}

\hybrid

\def\baselinestretch{1.2}

\catcode`\@=11

\def\marginnote#1{}
\newcount\hour
\newcount\minute
\newtoks\amorpm
\hour=\time\divide\hour by60
\minute=\time{\multiply\hour by60 \global\advance\minute by-\hour}
\edef\standardtime{{\ifnum\hour<12 \global\amorpm={am}%
	\else\global\amorpm={pm}\advance\hour by-12 \fi
	\ifnum\hour=0 \hour=12 \fi
	\number\hour:\ifnum\minute<10 0\fi\number\minute\the\amorpm}}
\edef\militarytime{\number\hour:\ifnum\minute<10 0\fi\number\minute}

\def\draftlabel#1{{\@bsphack\if@filesw {\let\thepage\relax
   \xdef\@gtempa{\write\@auxout{\string
      \newlabel{#1}{{\@currentlabel}{\thepage}}}}}\@gtempa
   \if@nobreak \ifvmode\nobreak\fi\fi\fi\@esphack}
	\gdef\@eqnlabel{#1}}
\def\@eqnlabel{}
\def\@vacuum{}
\def\draftmarginnote#1{\marginpar{\raggedright\scriptsize\tt#1}}

\def\draft{\oddsidemargin -.5truein
	\def\@oddfoot{\sl preliminary draft \hfil
	\rm\thepage\hfil\sl\today\quad\militarytime}
	\let\@evenfoot\@oddfoot	\overfullrule 3pt
	\let\label=\draftlabel
	\let\marginnote=\draftmarginnote
   \def\@eqnnum{(\theequation)\rlap{\kern\marginparsep\tt\@eqnlabel}%
\global\let\@eqnlabel\@vacuum}  }

\def\preprint{\twocolumn\sloppy\flushbottom\parindent 2em
	\leftmargini 2em\leftmarginv .5em\leftmarginvi .5em
	\oddsidemargin -.5in	\evensidemargin -.5in
	\columnsep .4in	\footheight 0pt
	\textwidth 10.in	\topmargin  -.4in
	\headheight 12pt \topskip .4in
	\textheight 6.9in \footskip 0pt
	\def\@oddhead{\thepage\hfil\addtocounter{page}{1}\thepage}
	\let\@evenhead\@oddhead	\def\@oddfoot{}	\def\@evenfoot{} }

\catcode`@=12
\relax

\def\figcap{\section*{Figure Captions\markboth
	{FIGURECAPTIONS}{FIGURECAPTIONS}}\list
	{Figure \arabic{enumi}:\hfill}{\settowidth\labelwidth{Figure 999:}
	\leftmargin\labelwidth
	\advance\leftmargin\labelsep\usecounter{enumi}}}
 \relax
\def\tablecap{\section*{Table Captions\markboth
	{TABLECAPTIONS}{TABLECAPTIONS}}\list
	{Table \arabic{enumi}:\hfill}{\settowidth\labelwidth{Table 999:}
	\leftmargin\labelwidth
	\advance\leftmargin\labelsep\usecounter{enumi}}}
 \relax
\def\reflist{\section*{References\markboth
	{REFLIST}{REFLIST}}\list
	{[\arabic{enumi}]\hfill}{\settowidth\labelwidth{[999]}
	\leftmargin\labelwidth
	\advance\leftmargin\labelsep\usecounter{enumi}}}
 \relax
\makeatletter
\newcounter{pubctr}
\def\publist{\@ifnextchar[{\@publist}{\@@publist}}
\def\@publist[#1]{\list
	{[\arabic{pubctr}]\hfill}{\settowidth\labelwidth{[999]}
	\leftmargin\labelwidth
	\advance\leftmargin\labelsep
	\@nmbrlisttrue\def\@listctr{pubctr}
	\setcounter{pubctr}{#1}\addtocounter{pubctr}{-1}}}
\def\@@publist{\list
	{[\arabic{pubctr}]\hfill}{\settowidth\labelwidth{[999]}
	\leftmargin\labelwidth
	\advance\leftmargin\labelsep
	\@nmbrlisttrue\def\@listctr{pubctr}}}
 \relax
\makeatother
\newskip\humongous \humongous=0pt plus 1000pt minus 1000pt

\newif\ifdtup

\relax

\def\s{\sigma}

\def\be{\begin{equation}}
\def\ee{\end{equation}}
\def\ba{\begin{eqnarray}}
\def\ea{\end{eqnarray}}

\def\S{\Sigma}

\begin{document}
\renewcommand{\theequation}{\arabic{equation}}
\newcommand{\beq}{\begin{equation}}
\newcommand{\eeq}[1]{\label{#1}\end{equation}}
\newcommand{\ber}{\begin{eqnarray}}
\newcommand{\eer}[1]{\label{#1}\end{eqnarray}}
\begin{titlepage}
\begin{center}

\hfill CERN--TH.7046/93\\
\hfill hep-th/9310121\\

\vskip .5in

{\large \bf $W_{\infty}$ SYMMETRY OF THE NAMBU--GOTO STRING\\
IN 4 DIMENSIONS}\\

\vskip 1.3in

{\bf Ioannis Bakas}
\footnote{Permanent Address: Department of Physics, University of
Crete,
GR--71409 Heraklion, Greece}
\footnote{e-mail address: BAKAS@SURYA11.CERN.CH}\\
\vskip .1in

{\em Theory Division\\
     CERN\\
     CH--1211 Geneva 23\\
     Switzerland}\\

\vskip .1in

\end{center}

\vskip 1in

\begin{center} {\bf ABSTRACT } \end{center}
\begin{quotation}
\noindent
 We consider a bosonic string propagating in 4--dim Minkowski
 space. We show that in the orthonormal gauge the classical system
exhibits
 a hidden $W_{\infty}$ chiral symmetry, arising from the equivalence
 of its transverse modes with the
 $SU(2)/U(1)$ coset model defined on the string world--sheet.
 Generalizations to other string backgrounds are proposed. We
 also define a Liouville--like transformation that maps solutions of
the
 $SU(2)/U(1)$ coset model into the solution space of two decoupled
 Liouville theories. Inverting this transformation, however, remains
an
 open problem.

\end{quotation}
\vskip1.0cm
CERN--TH.7046/93 \\
October 1993\\
\end{titlepage}
\vfill
\eject
\def\baselinestretch{1.2}
\baselineskip 16 pt
\noindent
Searching for hidden symmetries in string theory is a challenge that
has
not been met successfully yet.
The main difficulty is that there is no systematic way
to approach this problem from first principles. In this paper we
propose
a geometrical method that has the ambition to associate an integrable
system of non--linear differential equations to a given string
background.
Then, the conservation laws of this integrable system can be used to
construct an infinite set of symmetries for the corresponding string
model,
at least classically.

The main idea comes from an old work by Lund and Regge on the
relativistic
theory of vortices in superfluids [1]. Using their geometric
framework,
we examine at the classical level the presence of infinite symmetries
in
simple string models. Specializing to the case of a free bosonic
string
propagating in 4--dim Minkowski space, we find that in the
orthonormal
gauge the system exhibits a $W_{\infty}$ symmetry, one for the left
movers
and one for the right movers. The origin of this symmetry, which is
non--intuitive from the string point of view, is rather simple to
understand within the Lund--Regge formalism. There is an embedding
problem
in geometry associated with the evolution of this string, for which
the
dynamics of the metric $g$ and the extrinsic curvature $K$ is
determined
entirely by the classical equations of motion of the $SU(2)/U(1)$
2--dim
coset model defined on the string world--sheet.
Consequently, the $W_{\infty}$ symmetry of the latter is also
a symmetry of the 4--dim Nambu--Goto string, realized on the data
$(g,K)$.

Although this result is rather specific to the bosonic string
propagating
in 4--dim Minkowski space, it can be generalized to higher
dimensions.
Also, it can be generalized to include strings with
self--interactions.
In a separate publication we consider the example of a Nambu--Goto
string
in 4--dim Minkowski space
with Kalb--Ramond self--interaction of vortex type.
This model, which is actually the main subject of the Lund--Regge
theory,
corresponds to an integrable system obtained by perturbation of the
$SU(2)/U(1)$ coset model with the first thermal operator of $Z_{N}$
parafermions in the large $N$ limit [2].

We believe that appropriate generalization of the Lund--Regge
formalism
to a large variety of string backgrounds will produce further
relations
between the geometry of string dynamics and that of 2--dim integrable
systems
living on the string world--sheet.
          Inclusion of black--hole backgrounds with Minkowski
signature is also an interesting problem in this context. Then,
duality
symmetries between different string backgrounds might find a natural
interpretation in the theory of integrable systems. These are
problems
left for future work, while testing the limitations of our program in
search
of hidden symmetries in string theory.

We consider the Nambu--Goto string propagating in 4--dim Minkowski
space
with string variables $X_{\mu}(\sigma,\tau)$, $\mu = 0,~ 1,~ 2,~ 3$
and
signature $-+++$. The classical equations of
motion in the orthonormal gauge and in the Lorentz frame
$X_{0} = \tau$ are (see for instance [3])
\be
 ({\partial_{\sigma}}^{2} - {\partial_{\tau}}^{2}) X_{i} = 0 ~, ~~~~
 i= 1,~2,~3,
\ee
subject to the constraints
\be
 {(\partial_{\sigma} X_{i})}^{2} + {(\partial_{\tau} X_{i})}^{2} = 1
{}~,~~~~
   (\partial_{\sigma} X_{i}) (\partial_{\tau} X_{i}) = 0.
\ee
The results we describe in the sequel turn out to be independent of
the
choice of Lorentz frame in the $(\s , \tau)$ space, as we will see
later.

The starting point of the Lund--Regge formalism is the projection of
the
string world--sheet $\Sigma$, parametrized by $(\sigma , \tau)$,
on the $X_{0} = \tau$
hyperplane. The resulting 2-dim surface $S$ has Euclidean signature
with
metric given by
\be
 {ds}^{2} = {(\partial_{\sigma} X_{i})}^{2} {d \sigma}^{2} +
   {(\partial_{\tau} X_{i})}^{2} {d \tau}^{2}.
\ee
Clearly, it is sufficient to consider $S$ to know $X_{i}$ for $X_{0}
=
\tau$.
To satisfy the first constraint (2) we introduce an angle $\theta$
such
that
\be
 {(\partial_{\sigma} X_{i})}^{2} = cos^{2}\theta ~,~~~~
   {(\partial_{\tau} X_{i})}^{2} = sin^{2}\theta.
\ee
Then, for $X_{0} = \tau$, the evolution of the string is equivalent
to the
embedding of the 2--dim surface $S$ in 3--dim Euclidean space,
provided
that eq.(1) is also satisfied.

The formulation of an embedding problem requires apart from the
metric $g$,
(3), knowledge of the extrinsic curvature tensor $K$. Its components
are defined to be
\be
 K_{\sigma \sigma} = ({\partial_{\sigma}}^{2} X_{i}) Z_{i}^{(3)} ~ ,~
{}~
   K_{\tau \tau} = ({\partial_{\tau}}^{2} X_{i}) Z_{i}^{(3)} ~ , ~~
 K_{\sigma \tau} = K_{\tau \sigma} =
   (\partial_{\sigma} \partial_{\tau} X_{i}) Z_{i}^{(3)},
\ee
where $ Z^{(3)} = Z^{(1)} \times Z^{(2)} $ with
\be
 Z_{i}^{(1)} = {1 \over {cos\theta}} ~ \partial_{\sigma} X_{i} ~ ,
{}~~~~
   Z_{i}^{(2)} = {1 \over {sin\theta}} ~ \partial_{\tau} X_{i}.
\ee
In terms of these variables, the classical equations of motion (1)
are
equivalently stated as
\be
 K_{\sigma \sigma} - K_{\tau \tau} = 0.
\ee

Consider now the Gauss--Weingarten conditions for the embedding of a
2--dim
surface $S$ with data $(g,K)$ in 3--dim
Euclidean space (see for instance [4]).
For this to be at all possible, the following linear system of
equations
must be compatible
\be
\partial_{\sigma} \left( \begin{array}{c}
              Z^{(1)}\\
{}~\\
              Z^{(2)}\\
{}~\\
              Z^{(3)}\\
                         \end{array}   \right) =
\left( \begin{array}{ccc}
0               & \partial_{\tau}\theta    & K_{\s\s}/ cos\theta\\
{}~ & ~ & ~\\
-\partial_{\tau}\theta & 0        & K_{\s\tau}/ sin\theta\\
{}~ & ~ & ~\\
-K_{\s\s}/ cos\theta & -K_{\s\tau}/ sin\theta & 0\\
                                         \end{array}     \right)
\left( \begin{array}{c}
            Z^{(1)}\\
{}~\\
            Z^{(2)}\\
{}~\\
            Z^{(3)}\\
       \end{array}   \right),
\ee
\vskip 0.1in
\be
\partial_{\tau} \left( \begin{array}{c}
              Z^{(1)}\\
{}~\\
              Z^{(2)}\\
{}~\\
              Z^{(3)}\\
                      \end{array}    \right) =
\left( \begin{array}{ccc}
0       & \partial_{\s}\theta  &  K_{\s\tau}/ cos\theta\\
{}~ & ~ & ~\\
-\partial_{\s}\theta  & 0 & K_{\tau\tau}/ sin\theta\\
{}~ & ~ & ~\\
-K_{\s\tau}/ cos\theta & -K_{\tau\tau}/ sin\theta & 0\\
                                      \end{array}     \right)
\left( \begin{array}{c}
             Z^{(1)}\\
{}~\\
             Z^{(2)}\\
{}~\\
             Z^{(3)}\\
                  \end{array}    \right).
\ee

The resulting zero curvature condition yields a non--linear system of
second
order differential equations, known as Gauss--Codazzi equations.
It is convenient to introduce at this point
light--cone variables $(z,\bar{z})$ for the string, so that
$\partial = \partial_{\s} + \partial_{\tau}$, $\bar{\partial} =
\partial_{\s} - \partial_{\tau}$. We also introduce complex variables
$u(z, \bar{z})$, $\bar{u}(z, \bar{z})$ so that
\be
cos^2\theta = {\mid u \mid}^2 ~ , ~~~~ sin^2 \theta = 1 - {\mid u
\mid}^2
\ee
and
\be
K_{\s\s} + K_{\s\tau} =~ i ~ {u\partial\bar{u} - \bar{u}\partial u
\over
2 ~ \mid u \mid \sqrt{1- {\mid u \mid}^2}},
\ee
\be
K_{\s\s} - K_{\s\tau} =~ i ~{u \bar{\partial}\bar{u} -
\bar{u}\bar{\partial}u
\over 2 ~ \mid u \mid \sqrt{1 - {\mid u \mid}^2}}.
\ee
This parametrization of the $(g, K)$ data (8), (9) is compatible with
the
zero curvature condition we have, provided that eq.(7) is satisfied.
Then,
the resulting equations are
\be
\partial \bar{\partial} u + {\bar{u} \partial u \bar{\partial} u
\over
1- {\mid u \mid}^2} = 0,
\ee
\be
\partial \bar{\partial} \bar{u} + {u \partial \bar{u} \bar{\partial}
\bar{u} \over 1 - {\mid u \mid}^2} = 0.
\ee

It is straightforward to realize that in the present framework the
classical
equations of motion of the Nambu--Goto string in 4--dim Minkowski
space, in
the Lorentz frame $ X_{0} = \tau $, are equivalent to those of the
2--dim
coset model
\be
S = \int {dz d\bar{z} \over 1 - {\mid u \mid}^2}
(\partial u \bar{\partial} \bar{u} + \partial \bar{u} \bar{\partial}
u),
\ee
defined on the string world--sheet $\S$, with ${\mid u \mid}^2 \leq
1$.
This is precisely the action that arises classically in the
Lagrangian
description of the $SU(2)/U(1)$ coset model (defined on 2--dim
Minkowski
space) and exhibits curvature
singularities at the rim $ {\mid u \mid}^2 = 1 $;
its topology is not that of a round sphere, but of two bells
touching each other at the rim [5, 6]. Consequently, we may regard
the
transverse modes of the 4--dim
free bosonic string and the $SU(2)/U(1)$ coset model on $\S$ as being
classically equivalent, after having solved the
constraints (2).
The projection $S$ of $\S$ into the 3--dim Euclidean
space was an auxiliary geometrical
problem that helped us arrive at this result.

A natural question that arises at this point is whether the
equivalence
we have between the two theories depends on the special frame $ X_{0}
=
\tau $ or it is more general for projections of the string
world--sheet
on 3--dim slices related to each other by Lorentz transformations
in $(\s , \tau)$. As
has been pointed out in [1], the result is independent of the Lorentz
frame where the projection is performed. The reason is that choosing
a frame other than $ X_{0} = \tau $, say $ X_{0} = \tilde{\tau} $,
where
\be
\tilde{\tau} = sinh \alpha ~ \s + cosh \alpha ~ \tau ~ , ~~~~
\tilde{\s} = cosh \alpha ~ \s + sinh \alpha ~ \tau ,
\ee
also amounts to an embedding problem,
but this effectively introduces a spectral parameter in the problem
(better seen in the spin--1/2 representation of the Gauss--Weingarten
conditions) that depends on the Lorentz
transformation (16) into the new frame.
Since the non--linear equations that
follow from the zero curvature condition
are independent of the spectral parameter (as is the case for all
integrable systems), the equivalence between the two theories we
obtain
is independent of the Lorentz frame as well.

The solution of the Gauss--Weingarten system provides us with the
normal
and tangent unit vectors to the surface $S$, thus actually
constructing it.
When a spectral parameter is introduced, we obtain a family of
Euclidean
surfaces $S$ depending on it. However, they all correspond to the
same
string world--sheet $\S$ projected in different 3--slices.
Having presented the essential ingredients of the Lund--Regge
formalism
applicable to our string background, we discuss next the infinite
many
symmetries of the $SU(2)/U(1)$ coset model  and their realization on
the
$(g, K)$ data for string propagation. In a sense, the field variables
$\psi_{\pm}$ we adopt in the sequel provide us with a parafermionic
construction of $S$.

We first note that the classical equations of motion of the coset
model
(15) can be also expressed as a zero curvature condition for the
following
linear system
\be
\partial \Phi = \left( \begin{array}{cc}
0   & \psi_{+}\\
\\
-\psi_{-}  & 0\\
           \end{array}     \right)    \Phi,
\ee
\vskip 0.1in
\be
\bar{\partial} \Phi = \left( \begin{array}{cc}
{\mid u \mid}^2 - 1/2  & u \sqrt{1-{\mid u \mid}^2} ~ V_{+}\\
\\
\bar{u} \sqrt{1-{\mid u \mid}^2} ~ V_{-}  & - {\mid u \mid}^2 + 1/2\\
               \end{array}     \right)     \Phi.
\ee
This should be regarded as
an alternative formulation of the spin--1/2  representation of the
Gauss--Weingarten conditions (8), (9).
The variables $\psi_{\pm}$ in eqs.(17), (18) are taken to be
\be
\psi_{+} = {\partial u \over \sqrt{1-{\mid u \mid}^2}} ~ V_{+} ~ ,
{}~~~~
\psi_{-} = {\partial \bar{u} \over \sqrt{1-{\mid u \mid}^2}} ~ V_{-},
\ee
where
\be
V_{\pm} = exp \left\{ \pm {1 \over 2} \int dz {u \partial \bar{u} -
\bar{u} \partial u \over 1 - {\mid u \mid}^2} -
d \bar{z} {u \bar{\partial} \bar{u} - \bar{u} \bar{\partial} u \over
1-{\mid u \mid}^2} \right\}.
\ee
This way, we have naturally introduced the parafermions $\psi_{1} =
\psi_{+}$ and ${\psi_{1}}^{\dagger} = \psi_{-}$ of the $SU(2)/U(1)$
coset
model [5] by considering the zero curvature formulation
of its classical equations of motion.
The parafermion currents are chirally conserved on--shell, i.e.,
\be
\bar{\partial} \psi_{+} = 0 = \bar{\partial} \psi_{-}.
\ee
We also note for completeness that the
off--diagonal matrix elements in eq.(18) can be written as
$u \sqrt{1-{\mid u \mid}^2}~ V_{+} = \int dz (1-2{\mid u \mid}^2)
\psi_{+}$
and $\bar{u} \sqrt{1-{\mid u \mid}^2}~ V_{-} =
\int dz (1-2{\mid u \mid}^2) \psi_{-}$.

In a follow up paper [2], a suitable form of the linear system (17),
(18)
(and its perturbations),
with spectral parameter, is used further
to derive the local conservation laws of
the theory by applying the abelianization method of gauge connections
[7]. For the present model, all this method yields is what we
already know and nothing more. Namely, the (non--chiral) conservation
law of the $U(1)$ current associated with the rotational
symmetry $ u \rightarrow u e^{i \epsilon}$, $ \bar{u} \rightarrow
\bar{u} e^{-i \epsilon}$,
\be
\bar{\partial} J + \partial \bar{J} = 0,
\ee
where
\be
J = {u \partial \bar{u} - \bar{u} \partial u \over 1-{\mid u \mid}^2}
{}~ ,
{}~~~~ \bar{J} = {u \bar{\partial} \bar{u} - \bar{u} \bar{\partial} u
\over 1-{\mid u \mid}^2},
\ee
as well as an infinite set of chiral conservation laws that generate
the
$W_{\infty}$ algebra. Given the particular form of the associated
linear problem (17), (18), it is no surprise that the $W_{\infty}$
generators are bilinear in $\psi_{\pm}$. Explicitly we have in a
quasi--primary basis [8, 9]
\be
W_{s}(z) = \sum_{k=0}^{s-2} {{(-1)}^{s-k} \over s-1}
{s-1 \choose k+1} {s-1 \choose s-k-1} {\partial}^k \psi_{+}
{\partial}^{s-k-2} \psi_{-},
\ee
$s = 2, ~ 3, ~ 4, ~ \dots ~ $, up to an overall normalization
constant.

The variations of $u$ and $\bar{u}$ that generate the $W_{s}(z)$
currents
of the theory (15) have been constructed in [9]. The first few are
given by
\be
\delta_{\epsilon}^{(2)} u = \epsilon \partial u,
\ee
\be
\delta_{\epsilon}^{(3)} u = 2 \epsilon \left({\partial}^2 u +
2 {u \partial u \partial \bar{u} \over 1-{\mid u \mid}^2} \right) +
{\epsilon}^{\prime} \partial u,
\ee
\be
\delta_{\epsilon}^{(4)} u = 5 \epsilon \left({\partial}^3 u +
{2 (\partial u)^2 \partial \bar{u} + 3 u {\partial}^2 u \partial
\bar{u}
\over 1-{\mid u \mid}^2} +
3 {u^2 \partial u (\partial \bar{u})^2 \over {(1-{\mid u
\mid}^2)}^2}\right)
+ 5 {\epsilon}^{\prime} \left({\partial}^2 u + {u \partial u \partial
\bar{u}
\over 1-{\mid u \mid}^2} \right) + {\epsilon}^{\prime \prime}
\partial u,
\ee
etc, while for $\bar{u}$ we replace $u \leftrightarrow \bar{u}$ and
pick
up an overall $(-1)^s$ sign in the corresponding variation.
The commutation relations of the $\delta_{\epsilon}^{(s)}$
variations satisfy the centerless $W_{\infty}$ algebra. Explicitly
we have
\be
[\delta_{\epsilon_{1}}^{(2)} , \delta_{\epsilon_{2}}^{(2)}] =
\delta_{ \epsilon_{1}^{\prime} \epsilon_{2} - \epsilon_{1}
\epsilon_{2}^{\prime}}^{(2)},
\ee
\be
[\delta_{\epsilon_{1}}^{(2)} , \delta_{\epsilon_{2}}^{(3)}] =
\delta_{2 \epsilon_{1}^{\prime} \epsilon_{2} - \epsilon_{1}
\epsilon_{2}^{\prime}}^{(3)},
\ee
\be
[\delta_{\epsilon_{1}}^{(2)} , \delta_{\epsilon_{2}}^{(4)}] =
\delta_{3 \epsilon_{1}^{\prime} \epsilon_{2} - \epsilon_{1}
\epsilon_{2}^{\prime}}^{(4)} + {32 \over 5} ~
\delta_{\epsilon_{1}^{\prime \prime \prime} \epsilon_{2}}^{(2)},
\ee
\be
[\delta_{\epsilon_{1}}^{(3)} , \delta_{\epsilon_{2}}^{(3)}] =
2 ~ \delta_{\epsilon_{1}^{\prime} \epsilon_{2} - \epsilon_{1}
\epsilon_{2}^{\prime}}^{(4)} + {4 \over 5} ~
\delta_{2(\epsilon_{1}^{\prime \prime \prime} \epsilon_{2} -
\epsilon_{1}
\epsilon_{2}^{\prime \prime \prime}) + 3(\epsilon_{1}^{\prime}
\epsilon_{2}^{\prime \prime} - \epsilon_{1}^{\prime \prime}
\epsilon_{2}^{\prime})}^{(2)} ~ ,
\ee
etc\footnote{This also corrects a typographical mistake in [9].}.
Here we
have rescaled $W_{s}(z)$ with $2^{s-3} s! / (2s - 3)!!$.
The $SU(2)/U(1)$ coset model is formulated in Minkowski space and
the $W_{\infty}$ generators (24) are not real for odd spin $s$.
{}From now on, to achieve reality for all $s$, we will also
rescale $W_{s}(z)$ with
${i}^{s-2}$.

The embedding
problem of the surface $S$ in 3--dim Euclidean space,
with $K_{\s \s} = K_{\tau \tau} $, has an infinite
number of conservation laws associated with it.
In particular, using eqs.(10)--(12)
and the defining equations (19), (20) for the parafermions, we may
rewrire the $SU(2)/U(1)$ conservation laws in terms of the
geometrical data $(g , K)$. The $U(1)$ current yields the
non--chiral conservation law
\be
\bar{\partial} \left( \sqrt{{g_{\s \s} \over g_{\tau \tau}}} (K_{\s
\s} +
K_{\s \tau}) \right) + \partial \left( \sqrt{{g_{\s \s} \over
g_{\tau \tau}}} (K_{\s \s} - K_{\s \tau}) \right) = 0.
\ee
Taking into account the
rescaling introduced earlier and introducing the notation
\be
K_{\pm} = 2~ (K_{\s \s} \pm K_{\s \tau}) ~ , ~~~~
g_{-} = g_{\s \s} - g_{\tau \tau}
\ee
to simplify the expressions in what follows, we have for the
$W_{\infty}$
generators
\be
W_{2}(z) = {1 \over 4} ~ K_{+}^{2} +
{({\partial g_{-})}^{2} \over 16 ~ det g},
\ee
\be
W_{3}(z)  =  {g_{-} \over 2 \sqrt{det g}} ~
K_{+}^{3}  -
{\partial g_{-} \over
2 \sqrt{det g}} ~ \partial K_{+} +
{g_{-}
{(\partial g_{-})}^{2} + 2 detg ~
{\partial}^{2} g_{-}
\over 4 {(\sqrt{det g})}^{3}} ~
K_{+},
\ee
\begin{eqnarray}
W_{4}(z)  & = & {{(g_{-})}^{2}
\over det g} ~
K_{+}^{4} +
4 {( \partial K_{+})}^{2} -
{4 \over 5} {\partial}^{2} (K_{+}^{2}) +
{g_{-} \over 4
{(det g)}^{2}}  \left(3 g_{-} {(\partial g_{-})}^{2} + 8 det g ~
{\partial}^{2} g_{-} \right)  K_{+}^{2}  \nonumber \\
& - &  {g_{-}
\over det g} ~ (\partial g_{-})
{}~ \partial (K_{+}^{2}) -
{3 g_{-}^{2} + 8 det g \over  80 ~
{(det g)}^{3}} ~ {(\partial g_{-})}^{4} +
{3 {({\partial}^{2} g_{-})}^{2}
- 2 (\partial g_{-})
({\partial}^{3} g_{-}) \over 5 ~ det g},
\end{eqnarray}
etc. With these normalizations we have in general
\be
W_{s}(z) = 2^{2(s-2)} {\left({g_{\s \s} - g_{\tau \tau} \over
\sqrt{det g}}
\right)}^{s-2}  {(K_{\s \s} + K_{\s \tau})}^{s} ~ + ~ \cdots ~ ,
\ee
but the subleading terms look even more
complicated for higher values of $s$.
They are, nevertheless, computable.
To introduce the dependence of these infinitely many currents on the
original string variables $ \{X_{i}\} $, we simply have to use the
defining relations (3), (5), (6) for the metric and extrinsic
curvature tensors.

We have arrived at the result that
the $W_{\infty}$ symmetry of the $SU(2)/U(1)$ coset model carries
over to string theory and it is represented on the geometrical data
$(g, K)$ that describes the embedding of $S$ in terms of string
variables.
On--shell, i.e. when $S$ is actually embedded in the 3--dim slice
$X_{0} = \tau$ (or any other Lorentz equivalent frame in $(\s ,
\tau)$),
we obtain
the chiral conservation laws
\be
\bar{\partial}W_{s}(z) = 0.
\ee
Embedding problems in geometry are always described by the
Gauss--Codazzi
zero curvature conditions and therefore
we expect that there will be conservation
laws for the geometrical quantities involved. However, it is quite
difficult to find them and even more so to
write them down explicitly, in general. In the present case
it is the $W_{\infty}$ algebra and its parafermionic realization (24)
that
provide the algorithm for constructing the (higher) integrals of the
embedding.
Of course, we have two copies of $W_{\infty}$, one for the left
movers and
one for the right movers. The generators $\bar{W}_{s}(\bar{z})$ of
the other
sector are obtained by replacing $ \partial \rightarrow
\bar{\partial} $,
in which case $ K_{\s \s} + K_{\s \tau} \rightarrow K_{\s \s} -
K_{\s \tau}$. For them we have similarly $\partial
\bar{W}_{s}(\bar{z})
= 0 $.

We may also consider the infinitesimal variations of the $(g, K)$
data
induced by the $W_{\infty}$ variations of $u$ and $\bar{u}$.
On--shell,
they describe implicitly (via the Gauss--Weingarten equations)
the action of $W_{\infty}$ on the space of
classical string configurations. Using the classical equations of
motion (13), (14), we obtain the following result for the first few
variations:

\noindent{For $s = 2$,}
\be
\delta_{\epsilon}^{(2)} g_{\s \s}
= - \delta_{\epsilon}^{(2)} g_{\tau \tau}
= {1 \over 2} \epsilon ~ \partial g_{-} ,
\ee
\be
\delta_{\epsilon}^{(2)} K_{+} =  \partial \left(
\epsilon ~ K_{+} \right) ~ , ~~~
\delta_{\epsilon}^{(2)} K_{-} = - \epsilon ~
{\bar{\partial} g_{-} \over 4 ~ det g}
{}~ K_{+},
\ee
\noindent{for $s=3$,}
\be
\delta_{\epsilon}^{(3)} g_{\s \s} = - \delta_{\epsilon}^{(3)} g_{\tau
\tau}
= - 4 \epsilon ~ \partial \left( \sqrt{detg} ~ K_{+}
\right) - 2 {\epsilon}^{\prime} \sqrt{detg} ~ K_{+} ,
\ee
\ba
\delta_{\epsilon}^{(3)} K_{+} & = & {1 \over \sqrt{det g}} \left( 3
{\epsilon}^{\prime} g_{-} K_{+}^{2} + 3 \epsilon ~ g_{-}
\partial(K_{+}^{2}) + \epsilon ~ {{(g_{-})}^{2} + 4 det g
\over 2 ~ det g} ~ (\partial g_{-}) ~ K_{+}^{2}  \right. \nonumber \\
   & + & \mbox{} \left.  {\epsilon}^{\prime}
{g_{-} \over  det g} ~ {(\partial g_{-})}^{2} +
2 \epsilon ~ {g_{-} \over det g}
{}~ (\partial g_{-}) ({\partial}^{2} g_{-})
+ \epsilon ~ {{(g_{-})}^{2} + det g \over 2 ~ {(det g)}^{2}}
{}~ {(\partial g_{-})}^{3}  \right. \nonumber \\
   & + & \mbox{}  \left. {\epsilon}^{\prime \prime} \partial g_{-} +
3 {\epsilon}^{\prime} {\partial}^{2} g_{-} + 2 \epsilon ~
{\partial}^{3}
g_{-} \right),
\ea
\be
\delta_{\epsilon}^{(3)} K_{-}  =  - {\bar{\partial} g_{-} \over
8 {(\sqrt{det g})}^{3}} \left( 4 \epsilon ~ \left(g_{-} K_{+}^{2} +
{g_{-} {(\partial g_{-})}^{2} \over 4 ~ det g} + {\partial}^{2} g_{-}
\right) + 2 {\epsilon}^{\prime} \partial g_{-} \right)
\ee
and so on. Despite their complicated nature, these are hidden
symmetries
of string propagation in the orthonormal gauge after imposing the
constraints (2). As for the $\bar{W}_{\infty}$ sector of the
symmetry,
similar expressions are obtained by computing the
$\delta_{\bar{\epsilon}}^{(s)}$ variations. Alternatively, one may
consider the bosonic string action and compute directly the
$W_{\infty}$
variations of $\{X_{i}\}$ associated with the conserved currents
(37).
This might lead to a better geometrical understanding of the
infinitely
many symmetries that govern string propagation.

It is fair to say that the idea to associate integrable systems of
non--linear differential equations to embedding problems has been
considered before in various occassions, but in a different context.
For example, it is well known that the sine--Gordon equation arises
as
a constant curvature condition of 2--dim Riemannian surfaces, in a
special coordinate frame (see for instance [1, 10]). Also, the
embedding of 2--dim surfaces into spaces equipped with a Lie algebra
structure is associated to Toda theories [11]. More recently, the
extrinsic geometry of 2--dim surfaces chirally embedded in $CP^{n}$
has
been used extensively to formulate various geometrical aspects of
classical
$W_{n}$--symmetries ([12] and references therein). In the present
context,
however, the embedding we are considering is prescribed by a
different
physical problem, namely the classical evolution of a bosonic string.
Hence, the corresponding integrable system, which is the
$SU(2)/U(1)$ coset model, is
of a different type than the ones considered before.

It would be interesting to find out whether more general coset models
of
(rational) conformal field theory (or their integrable perturbations)
could be associated with string propagation in arbitrary backgrounds,
via a Lund--Regge type formalism. A good starting point for further
generalization is to consider string backgrounds with a time--like
Killing symmetry. This way the interplay between strings, geometry
and
$W_{\infty}$--type algebras might be proven useful for looking at the
problem of hidden symmetries, in general.

There have also been a few remarkable attempts in the past to obtain
an infinite set of conservation laws for string dynamics. Some of
them [13]
have been influenced by the integrability aspects of non--linear
$\s$--models. Another group theoretical approach has been proposed by
Pohlmeyer [14] (see also [15]), where the loop equation of the
Nambu--Goto string has been interpreted as a collection of
representation
conditions for an infinite dimensional algebra of reparametrization
invariant charges. Also, Witten has made some interesting general
remarks
on the subject [16]. It would be useful to study the relation of our
results with theirs, since the geometric and algebraic methods
employed in each case are quite different. This way we hope to
sharpen our
understanding of the problem.

We turn now to the question of finding solutions to the classical
equations of motion (13), (14), since to any given solution there
corresponds a surface $S$ (and hence $\S$) actually built by solving
the Gauss--Weingarten system for its normal and tangent vectors.
This is a general question in the theory of gauged WZW models, since
the action (15) arises classically for $SU(2)$
after performing the neccessary
gauge field integrations. Although there is already some work on the
explicit
form of the classical solutions
of coset models [17], we will develop a different approach
emphasizing the role of Liouville theory in string propagation. We
also
think that uncovering the Liouville--like structure of the simplest
2--dim coset model $SU(2)/U(1)$
might give some new ideas about the more general situation.

An obvious set of solutions is given by
\be
\partial \bar{u} = 0 = \bar{\partial} u,
\ee
in analogy with the instanton solutions of the ordinary $O(3)$
non--linear
$\s$--model. In the present model, however, this class is not
previledged
because the volume of the target space is infinite and so is the
action.
To examine the solution space of the $SU(2)/U(1)$ coset model, in
general,
we introduce a transformation that maps its classical equations of
motion
to two decoupled Liouville equations. ``Half" of this transformation
has been considered before [18] in the study of 2--dim black--holes
and
Liouville theory in the first order formalism
\footnote{I thank E. Verlinde for some discussions on this
point.}. It also has a useful interpretation when applied to the
ordinary
$O(3)$ non--linear $\s$--model.

To be more precise we define two real fields $\varphi_{1}$ and
$\varphi_{2}$ as
\be
e^{\varphi_{1}} = {\partial u \bar{\partial} \bar{u} \over
{(1-{\mid u \mid}^2)}^2} ~ , ~~~~
e^{\varphi_{2}} = {\partial \bar{u} \bar{\partial} u
\over {(1-{\mid u \mid}^2)}^2}.
\ee
Then, it is straightforward to verify that eqs.(13), (14) become
\be
\partial \bar{\partial} \varphi_{1} = 2 ~ e^{\varphi_{1}} ~ , ~~~~
\partial \bar{\partial} \varphi_{2} = 2 ~ e^{\varphi_{2}}.
\ee
If we had applied the same transformation to the equations of motion
of
the ordinary $\s$--model with target space metric $ h(u, \bar{u}) =
{(1-{\mid u \mid}^2)}^{-2}$, instead of ${(1-{\mid u \mid}^2)}^{-1}$,
we would have obtained the
$\widehat{SL}(2)$ affine Toda system
\be
\partial \bar{\partial} \varphi_{1} = 2 ~ e^{\varphi_{1}}  - 2 ~
                                             e^{\varphi_{2}} ~ , ~~~~
\partial \bar{\partial} \varphi_{2} = -2 ~ e^{\varphi_{1}}  +2 ~
                                             e^{\varphi_{2}}.
\ee
The latter result underlies the well known relation between the
$O(3)$ non--linear $\s$--model and the $sinh$--Gordon equation [19].
We also note for completeness that the transformation
we are using here yields
a closed system of differential equations for $\varphi_{1}$ and
$\varphi_{2}$ in the class of target space metrics $h(u, \bar{u}) =
{(1-{\mid u \mid}^2)}^{-\alpha}$, only for $\alpha = 1, ~ 2$.

The map between the solution space of the $SU(2)/U(1)$
coset model and that of two decoupled Liouville theories should be
understood further, in view of the relevance
they both have in string propagation.
It is clear that there is a $GL(2)$ redundance in associating
$\varphi_{1}$, $\varphi_{2}$ to $u$, $\bar{u}$ due to the
invariance of the transformation (45) under
\be
u \rightarrow {a u + b \over c u + d} ~ , ~~~~
\bar{u} \rightarrow {d \bar{u} + c \over b \bar{u} + a} ~ ,
 ~~~~ a d - b c \neq 0,
\ee
in exact analogy with Liouville theory. To preserve the condition
${\mid u \mid}^2 \leq 1$, introduced by solving the first constraint
(2), we have to restrict the group to ${GL}^{+}(2)$ with
non--negative
$a, ~ b, ~ c, ~ d$ and $a d - b c > 0$. We also have a rotational
invariance in eq.(45) under the global transformations $u \rightarrow
u e^{i \epsilon}$, $\bar{u} \rightarrow \bar{u} e^{-i \epsilon}$.
Modulo this redundance, the invertibility
of the Liouville--like transformation (45) should be investigated
in favor of expressing the general solution of the $SU(2)/U(1)$ coset
model in terms of two
Liouville fields. For the latter, the general solution is
known to be
\be
e^{\varphi_{1}} = {\partial F_{1} \bar{\partial} \bar{F}_{1} \over
{(1-F_{1} \bar{F}_{1})}^{2}}  ~ , ~~~~
e^{\varphi_{2}} =  {\partial F_{2} \bar{\partial} \bar{F}_{2} \over
{(1-F_{2} \bar{F}_{2})}^{2}},
\ee
where $ \bar{\partial} F_{i} = 0 $ and $ \partial \bar{F}_{i} = 0 $,
but otherwise arbitrary.
We note that for the ``instanton"--like solutions (44) all this is
rather trivial, since only $\varphi_{1}$ needs to be considered
with $F_{1} = u$ and $\bar{F}_{1} = \bar{u}$.
In the general case, however, the fact that $\varphi_{1}$
and $\varphi_{2}$ still satisfy the Liouville equations (46) is quite
remarkable.  The description of classical
string dynamics directly in terms of Liouville fields remains an open
problem.

A related issue in this context is to understand the description of
the $W_{\infty}$ symmetry in terms of
two Liouville fields. The only local currents
we know for them, which are functionally independent and
chirally conserved, are their corresponding stress--energy tensors.
The
stress--energy tensor $\psi_{+} \psi_{-} = \partial u \partial
\bar{u} /
(1-{\mid u \mid}^{2})$ of the $SU(2)/U(1)$ coset model can be easily
described in Liouville terms, but for the higher spin generators of
$W_{\infty}$
non--local conservation laws of the system (45) have to
be taken into account. Their systematic construction is beyond the
scope of the present work.

These questions as well as the possibility to associate
2--dim integrable systems of non--linear differential equations to
string
propagation in more general backgrounds are all under investigation.
As a byproduct, we hope to gain a better understanding of
the hidden symmetries
in string theory. Their relevance in quantum theory might be easier
to
examine within the framework of covariant quantization on the space
of
classical string solutions [20].
The issue of boundary conditions should also be
properly addressed in the future. Finally, the effect of adding
extrinsic
curvature terms to the Nambu--Goto action
{\em \`a la} Polyakov [21] needs to be
clarified within the present framework.

\vskip 1cm

\centerline{\bf Acknowledgments}
\vskip 0.1in
\noindent
I am grateful to E. Kiritsis for many useful discussions and
collaboration
on some of the issues considered here,
to E. Floratos for encouragement and
T. Tomaras for constructive critisism, all very much needed after my
17
month long military service. I also thank A. Polychronakos for a
critical
reading of the manuscript.

\vskip 1.5cm

\centerline{\bf REFERENCES}
\begin{enumerate}
\item F. Lund and T. Regge, Phys. Rev. \underline{D14} (1976) 1524.
\item I. Bakas, {\em ``Conservation Laws and Geometry of
Perturbed Coset Models"},
preprint CERN--TH.7047, October 1993.
\item M. Green, J. Schwartz and E. Witten, {\em ``Superstring
Theory"},
volume I, Cambridge University Press, Cambridge, 1987.
\item L. Eisenhart, {\em ``Riemannian Geometry"}, Princeton
University
Press, Princeton, New Jersey, 1964.
\item K. Bardacki, M. Crescimanno and E. Rabinovici, Nucl. Phys.
\underline{B344} (1990) 344; A. Giveon, Mod. Phys. Lett.
\underline{A6} (1991) 2843;
E. Kiritsis, Mod. Phys. Lett. \underline{A6}
(1991) 2871.
\item E. Witten, Phys. Rev. \underline{D44} (1991) 314.
\item V. Drinfeld and V. Sokolov, Jour. Sov. Math. \underline{30}
(1985)
1975; D. Olive and N. Turok, Nucl. Phys. \underline{B257(FS14)}
(1985) 277.
\item I. Bakas and E. Kiritsis, Nucl. Phys. \underline{B343} (1990)
185.
\item I. Bakas and E. Kiritsis, Phys. Lett. \underline{B301} (1993)
49.
\item S. Coleman, {\em ``Aspects of Symmetry: Selected Erice
Lectures"},
Cambridge University Press, Cambridge, 1985.
\item M. Saveliev, Theor. Math. Phys. \underline{60} (1984) 638;
A. Leznov and M. Saveliev, Acta Applic. Math. \underline{16} (1989)
1.
\item J.--L. Gervais and Y. Matsuo, Phys. Lett. \underline{B274}
(1992)
309; Comm. Math. Phys. \underline{152} (1993) 317.
\item A. Isaev, Sov. Phys. JETP Lett. \underline{33} (1981) 341;
A. Balachandran and A. Stern, Phys. Rev. \underline{D26} (1982) 1436.
\item K. Pohlmeyer, Phys. Lett. \underline{B119} (1982) 100; Comm.
Math.
Phys. \underline{105} (1986) 629, \underline{114}
(1988) 351.
\item K. Pohlmeyer and K.--H. Rehren, Comm. Math. Phys.
\underline{105}
(1986) 593, \underline{114} (1988) 55, \underline{114} (1988) 177.
\item E. Witten, Phil. Trans. Roy. Soc. Lond. \underline{A320} (1989)
349.
\item  I. Bars and K. Sfetsos, Mod. Phys. Lett. \underline{A7} (1992)
1091;
E. Kiritsis, private communication.
\item E. Martinec and S. Shatashvili, Nucl. Phys. \underline{B368}
(1992)
338.
\item K. Pohlmeyer, Comm. Math. Phys. \underline{46} (1976) 207.
\item E. Witten and C. Crnkovic, in {\em ``Three Hundred Years of
Gravitation"}, ed. by S. Hawking and W. Israel, Cambridge University
Press,
Cambridge, 1987; C. Crnkovic, Class. Quant. Grav. \underline{5}
(1988) 1557.
\item A. Polyakov, Nucl. Phys. \underline{B268} (1986) 406.
\end{enumerate}

\end{document}